\newif{\ifjournal}
  \journalfalse

\ifjournal
  \documentclass[]{aa}
  \usepackage{graphicx,times,amssymb,natbib}
\else
  \documentclass{paper}
\fi

\renewcommand{\d}{\mathrm{d}}

\begin{document}

\title{Arc statistics in cosmological models with dark energy}
\ifjournal
  \author{Matthias Bartelmann\inst{1}, Massimo Meneghetti\inst{1,2},
    Francesca Perrotta\inst{3,4,5}, Carlo Baccigalupi\inst{4,5}, \and
    Lauro Moscardini\inst{6}}
  \institute
   {$^1$ Max-Planck-Institut f\"ur Astrophysik, P.O.~Box 1317,
    D--85741 Garching, Germany\\
    $^2$ Dipartimento di Astronomia, Universit\`a di Padova, vicolo
    dell'Osservatorio 2, I--35122 Padova, Italy\\
    $^3$ Osservatorio Astronomico di Padova, Vicolo dell'Osservatorio
    5, I--35122 Padova, Italy\\
    $^4$ Lawrence Berkeley National Laboratory, 1 Cyclotron Road,
    Berkeley, CA 94720, USA\\
    $^5$ SISSA/ISAS, Via Beirut 4, 34014 Trieste, Italy\\
    $^6$ Dipartimento di Astronomia, via Ranzani 1, I--40127 Bologna,
    Italy}
  \authorrunning{M. Bartelmann et al.}
  \titlerunning{Arc Statistics in Cosmological Models with Dark Energy}
\else
  \author{Matthias Bartelmann$^1$, Massimo Meneghetti$^{1,2}$,
    Francesca Perrotta$^{3,4,5}$, Carlo Baccigalupi$^{4,5}$,\\
    Lauro Moscardini$^6$\\
    $^1$ Max-Planck-Institut f\"ur Astrophysik, P.O.~Box 1317,
    D--85741 Garching, Germany\\
    $^2$ Dipartimento di Astronomia, Universit\`a di Padova, vicolo
    dell'Osservatorio 2, I--35122 Padova, Italy\\
    $^3$ Osservatorio Astronomico di Padova, Vicolo dell'Osservatorio
    5, I--35122 Padova, Italy\\
    $^4$ Lawrence Berkeley National Laboratory, 1 Cyclotron Road,
    Berkeley, CA 94720, USA\\
    $^5$ SISSA/ISAS, Via Beirut 4, 34014 Trieste, Italy\\
    $^6$ Dipartimento di Astronomia, via Ranzani 1, I--40127 Bologna,
    Italy}
\fi

\date{{\em Astronomy \& Astrophysics, in press}}

\newcommand{\abstext}
  {We investigate how the probability of the formation of giant arcs
   in galaxy clusters is expected to change in cosmological models
   dominated by dark energy with an equation of state $p=w\rho c^2$
   compared to cosmological-constant or open models. To do so, we use
   a simple analytic model for arc cross sections based on the
   Navarro-Frenk-White density profile which we demonstrate reproduces
   essential features of numerically determined arc cross
   sections. Since analytic lens models are known to be inadequate for
   accurate absolute quantifications of arc probabilities, we use them
   only for studying changes relative to cosmological-constant
   models. Our main results are (1) the order of magnitude difference
   between the arc probabilities in low density, spatially flat and
   open CDM models found numerically is reproduced by our analytic
   model, and (2) dark-energy cosmologies with $w>-1$ increase the arc
   optical depth by at most a factor of two and are thus unlikely to
   reconcile arc statistics with spatially flat cosmological models
   with low matter density.}

\ifjournal
  \abstract{\abstext%
    }
\else
  \begin{abstract}
    \abstext
  \end{abstract}
\fi

\maketitle

\section{Introduction}

The statistics of giant luminous arcs in the cores of galaxy clusters
has long been recognised as a potentially powerful cosmological probe
(e.g.~\citealt{WU96.1,BA98.2}). Arcs are formed by gravitational
lensing from sources which happen to lie close to the caustic curves
of a cluster lens, where magnification and distortion are particularly
strong.

The cosmological power of arc statistics derives from at least two,
maybe three principal reasons. First, for clusters to be efficient
lenses, they have to be located approximately half-way between the
sources, typically around redshift unity, and the observer. Depending
mostly on the mean cosmic matter density, parameterised by $\Omega_0$,
clusters form earlier or later in cosmic history if the matter density
is low or high, respectively. In high-density model universes, the
cluster population at the redshifts mostly relevant for lensing,
$z\sim0.3-0.4$ is substantially less rich than in low-density
universes, reducing the number of available efficient lenses
dramatically (e.g.~\citealt{RI92.1,BA93.1,LA93.1,LA94.1}).

The second principal reason is that strong lensing is a highly
nonlinear phenomenon in the sense that it requires the lensing mass
distribution to be supercritical for strong lensing, which means that
a suitable combination of surface mass density and gravitational tidal
field needs to be large enough, and that, once a lens is
supercritical, even small changes in both can change significantly the
length of the caustic curves, and thus the lens' ability for strong
lensing.

Different cosmological models predict the mass distribution in
clusters to be more or less concentrated. Numerical simulations
consistently show that, the earlier a dark-matter halo forms, the more
concentrated it is because it appears to keep a record of the mean
cosmic density at the time when it formed \citep{NA96.1,NA97.1}.
Structure forms later in spatially flat than in open, low density
cosmological models, thus halos in models with cosmological constant
are generally less concentrated than halos in open models.

A possible third reason is that the gravitational tidal field at the
location of the lens plays a very important role
\citep{BA95.1,ME01.1}. It is strong if lenses are highly asymmetric,
as clusters frequently are, and if the surrounding matter distribution
is highly irregular. It is possible that, if cosmic structure forms
later, cluster mass distributions are less relaxed and thus more
asymmetric, and that also the degree of irregularity in their
neighbourhood is different than if structures formed earlier. On the
other hand, clusters forming earlier are built from subhalos which
tend to be more concentrated and thus more strongly gravitationally
bound, hence substructures could then persist within clusters for a
longer time and contribute to the asymmetry.

Unfortunately, the combination of these effects renders analytic
models for arc statistics entirely inadequate for accurate
quantitative predictions of arc probabilities \citep{ME03.1}. The
effects of cosmology on cluster compactness and asymmetry, and on the
tidal field of the matter surrounding the clusters, cannot be captured
by reasonably simple analytic lens models. Numerical simulations of
arc statistics, using clusters formed in sufficiently large $N$-body
simulations as lenses, led to the surprising result that the expected
number of giant luminous arcs on the sky differs by orders of
magnitude between different cosmological models. While a model with
critical matter density and no cosmological constant fell below the
observed number of arcs, extrapolated to the full sky, by two orders
of magnitude, a $\Lambda$CDM model with $\Omega_0=0.3$ failed by one
order of magnitude, and only a low-density open model with
$\Omega_0=0.3$ produced approximately the right number of arcs
\citep{BA98.2}.

The statistics of quasars multiply imaged by galaxies has often been
used for constraining cosmological parameters. The basic argument is
that the number of lenses and their redshifts should increase as
$\Omega_\Lambda$ increases, which typically yields upper limits on
$\Omega_\Lambda\lesssim0.6-0.7$ \citep{KO96.1,FA98.1,QU99.1}, although
discrepant results have also been found
\citep{CH99.1,HE99.1,KE02.1}. We emphasise that the sensitivity of
cluster lensing to $\Omega_\Lambda$ is of a different nature. Since
clusters form much later in cosmic history than galaxies, the volume
effect is negligible, but $\Omega_\Lambda$ changes the dynamics of
cluster formation and thus their core structure, to which strong
lensing is highly sensitive.

These numerical results of \citet{BA98.2} were tested by
\citet{CO99.2} and \citet{KA00.1} using analytic models based on
singular isothermal spheres. They could confirm the sensitivity of arc
statistics to $\Omega_0$, but found only a very weak dependence on
$\Omega_\Lambda$, in contrast to the numerical results. The isothermal
sphere has two disadvantages with respect to arc statistics. First,
arc cross sections are very sensitive to asymmetries in the cluster
mass distribution, thus axially symmetric models lack a property which
is crucially important for arc statistics. Second, numerical
simulations show that the central density concentration of clusters
depends on cosmology, and this potentially important feature is not
reproduced by the scale-free isothermal models either. We shall
construct in this paper an analytic model which qualitatively
reproduces the earlier numerical results.

\citet{BA98.2} used two completely different types of $N$-body codes
for simulating galaxy clusters. Numerous subsequent tests of the
results showed that the arc numbers derived could be off by factors of
perhaps $1.5$ to $2$, but that there was no way how order-of-magnitude
differences could be bridged \citep{ME00.1,FL00.1}. The problem became
substantially more acute when measurements of the cosmic microwave
background (CMB), combined with observations of supernovae of type Ia
and large-scale galaxy surveys, left very little room for model
universes which are not spatially flat and have density parameters
much different from $\Omega_0\approx0.3$
(e.g.~\citealt{RI98.1,PE99.1,LE01.1,AB02.1,EF02.1,NE02.1,WA02.1,HA02.1}).
Obviously, there is an interesting discrepancy between the statistics
of arcs seen on the sky, and the probability for arcs produced in
cosmological models which are convincingly required by various other
observations. While observations consistently indicate a high
probability for arc formation in clusters
\citep{LE94.1,GI94.1,LU99.1,ZA03.1,GL03.1}, we should point out that
the discrepancy between theory and observations is so far only based
on a single set of simulated clusters.

A spatially flat universe with low matter density and a cosmological
constant is extremely difficult to justify theoretically. The vacuum
energy density provided by the cosmological constant is tens of orders
of magnitudes below any natural scale which is conceivable in particle
physics (see \citealt{CA01.1} for a review on the cosmological
constant problem). This difficulty motivated the introduction of a
more general concept for a vacuum energy cosmological component, now
widely known as dark energy. The theoretical and observational aspects
of the dark energy are one of the most important issues in modern
cosmology (see \citealt{PE03.1} for a review). In general, the most
important difference of a dark energy component compared to a
cosmological constant is that its equation of state, $w$, can be
different from $-1$, generally implying a time variation. It should be
noted that recent analyses of CMB data seem to favour a value of $w$
very close to $-1$, albeit these results suffer to some degree from
parameter degeneracies and are typically obtained under restrictive
assumptions \citep{DO03.1,ME03.3,SP03.1}.

Recently, \citet{BA02.1} argued that dark matter halos in simple dark
energy models should be more concentrated than in cosmological
constant models with the same dark energy density today.  The main
reason is that halos form earlier in dark energy models, allowing
them to be more compact. Several different recipes for describing halo
concentrations found in numerical simulations as a function of their
formation time lead to consistent results. The halo concentration
increases noticeably in the interval $-1\le w\le-0.6$. For higher
values of $w$, which are too high for the cosmic acceleration to agree
with recent data \citep{RI98.1,PE99.1}, there is a strongly opposing
effect related to the amplitude of fluctuations in the CMB: The high
level of the Integrated Sachs-Wolfe (ISW) effect on the large scale
CMB anisotropies leads to a sharp decrease in the normalisation of the
dark matter power spectrum \citep{BA02.1}. Within a cosmologically
interesting range for the equations of state of the dark energy, the
balance between the ISW effect and the earlier formation of halos is
quite delicate, but halos can typically be expected to be more
concentrated.

The ability of a galaxy cluster to produce giant arcs depends
sensitively on the concentration of its mass profile because of the
nonlinearity of the strong lensing effect. We therefore wish to
investigate how the probability for arc formation changes in dark
energy models, compared to cosmological-constant or open models. For
doing so, we use a simple, analytic description for the arc cross
section of a cluster of given mass, which we demonstrate to possess
the relevant features of the fully numerical results. Although it has
been shown that analytic models are inadequate for quantitatively
reliable arc statistics, we are here interested only in the relative
change of the arc-formation probability caused by changes in the
cosmological model.

Section~2 of the paper introduces the cosmological background model.
Section~3 describes our simple analytic model for the arc cross
section of a galaxy cluster. We then use this model in Sect.~4 for
calculating arc probabilities, and summarise our conclusions in
Sect.~5

\section{Cosmological Model}

Dark energy is characterised by a negative pressure, $p=w\rho c^2$,
where $\rho c^2$ is the mean energy density of the universe and the
equation of state $w$ assumes negative values in order to produce
cosmic acceleration according to the data from type Ia supernovae
\citep{RI98.1,PE99.1}. Theoretical models of dark energy, such as
Quintessence scalar fields, generally predict a time variation of the
equation of state, as well as the presence of dark energy fluctuations
on super-horizon cosmological scales (see e.g.~\citealt{PE03.1} and
references therein). In this work, we concentrate on the very basic
aspect of dark energy, by neglecting the spatial inhomogeneities and
assuming $w$ to be a constant. In this case, the adiabatic equation
requires the equivalent matter density $\rho_\mathrm{Q}$ of the dark
energy to change with the cosmological scale factor $a$ as
\begin{equation}
  \rho_\mathrm{Q}=\rho_\mathrm{Q,0}\,a^{-3(1+w)}\;,
\label{eq:Q1}
\end{equation}
starting from the density $\rho_\mathrm{Q,0}$ today. Obviously,
cosmological constant models are retained setting $w=-1$. Replacing
the conventional cosmological-constant term by a dark-energy term,
Friedmann's equation reads
\begin{equation}
  H^2(a)=H_0^2\,\left[
    \Omega_0\,a^{-3}+\Omega_\mathrm{Q}\,a^{-3(1+w)}
  \right]\;,
\label{eq:Q2}
\end{equation}
assuming with support from recent measurements of anisotropies in the
cosmic microwave background that the curvature term is
negligible. Here, $H(a)$ is the Hubble parameter as a function of $a$,
$H_0$ is the Hubble constant, and $\Omega_0$ is the density parameter
for non-relativistic matter.

The main consequences for the structure formation process of this
modified term in Friedmann's equation have been detailed in an earlier
paper \citep{BA02.1}, so we summarise them only briefly here. Starting
from a low-density cosmological constant model, keeping $\Omega_0$
fixed and increasing $w$, the cosmic volume per unit redshift shrinks,
and the linear growth factor for cosmic structures starts rising
earlier, hence structures start forming earlier if the cosmological
constant is replaced by a dark energy component. Additional effects
are that the parameters characterising halo formation, i.e.~the linear
overdensity $\delta_\mathrm{c}$ and the virial overdensity
$\Delta_\mathrm{v}$, are changed (cf.~\citealt{WA98.1,MA03.1}).

As we anticipated in the introduction, a further important consequence
of dark energy models is that the gravitational potential of the
density fluctuations changes more rapidly with time, leading to an
increased integrated Sachs-Wolfe effect on the CMB fluctuations at
large angular scales. Given the observed level of CMB power on large
angular scales\footnote{The only existing data are from the
Differential Microwave Radiometer (DMR) on board the COsmic Background
Explorer satellite (COBE, see \citealt{SM99.1} and references
therein); more data will be available in the near future from the
Microwave Anisotropy Probe (MAP, \texttt{map.gsfc.nasa.gov}) and the
\emph{Planck} (\texttt{astro.estec.esa.nl/SA-general/Projects/Planck})
satellites.}, a decreasing fraction of the observed anisotropies can
thus be attributed to the primordial CMB fluctuations, hence the
normalisation of the power spectrum has to be reduced as $w$ is
increased in order to be compliant with the COBE-DMR data. The main
result is that the power-spectrum normalisation $\sigma_8$ is
decreasing gently as $w$ is increased from $-1$ to $\sim-0.6$, and
then turns to drop more rapidly as the ISW effect intensifies. Here
and below, we adopt the CDM power spectrum with the
Harrison-Zel'dovich power-law index $n=1$ and the transfer function
given by \citet{BA86.1}. Throughout, we use $\Omega_0=0.3$ and
$\Omega_\mathrm{Q}=0.7$. It should be noted here that our results will
be sensitive to the exact value of $n$ as well as other cosmological
parameters.

\section{Arc Cross Sections}

\subsection{Halo Model}

We will assume in the following that the average radial density
profile of galaxy clusters can be described by the profile found in
numerical simulations by \citet[ hereafter NFW]{NA96.1},
\begin{equation}
  \rho(r)=\frac{\rho_\mathrm{s}}{r/r_\mathrm{s}(1+r/r_\mathrm{s})^2}\;,
\label{eq:1}
\end{equation}
where $\rho_\mathrm{s}$ is a density scale, and $r_\mathrm{s}$ a scale
radius. The ratio between $r_\mathrm{s}$ and the radius $r_{200}$
enclosing a mean halo density of 200 times the critical density is
called concentration, $c=r_\mathrm{s}/r_{200}$. The two parameters,
$\rho_\mathrm{s}$ and $r_\mathrm{s}$, are not independent. Numerical
simulations showed that the concentration parameter $c$ depends on the
halo mass, which is thus the only free parameter.

Numerically simulated halos tend to be the more concentrated the
earlier they form. Their central density apparently reflects the mean
cosmic density at the time of their formation. Since halos of higher
mass are formed later in hierarchical models than halos of lower
mass, the concentration is decreasing with halo mass. Based on these
findings, several algorithms were designed for computing halo
concentrations from halo masses.

The algorithm by \citet{NA97.1} first assigns to a halo of mass $M$ a
collapse redshift $z_\mathrm{coll}$ defined as the redshift at which
half of the final halo mass is contained in progenitors more massive
than a fraction $f_\mathrm{NFW}$ of the final mass. Then, the density
scale of the halo is assumed to be some factor $C$ times the mean
cosmic density at the collapse redshift. They recommend setting
$f_\mathrm{NFW}=0.01$ and $C=3\times10^3$ because their numerically
determined halo concentrations were well fit assuming these values.

\citet{BU01.1} suggested a somewhat simpler algorithm. Haloes are
assigned a collapse redshift defined such that the non-linear mass
scale at that redshift is a fraction $f_\mathrm{B}$ of the final halo
mass. The halo concentration is then assumed to be a factor $K$ times
the ratio of the scale factors at the redshift when the halo is
identified and at the collapse redshift. Comparing with numerical
simulations, they found $f_\mathrm{B}=0.01$ and $K=4$. This algorithm
reflects the change of halo concentrations with redshift more
accurately than the approach by \citet{NA97.1} predicts.

A third algorithm was suggested by \citet{EK01.1}. They assigned the
collapse redshift to a halo of mass $M$ by requiring that the suitably
defined amplitude of the linearly evolving power spectrum at the mass
scale $M$ equals a constant $C_\mathrm{ENS}^{-1}$. Numerical results
are well represented setting $C_\mathrm{ENS}=28$.

\begin{figure}[ht]
  \includegraphics[width=\hsize]{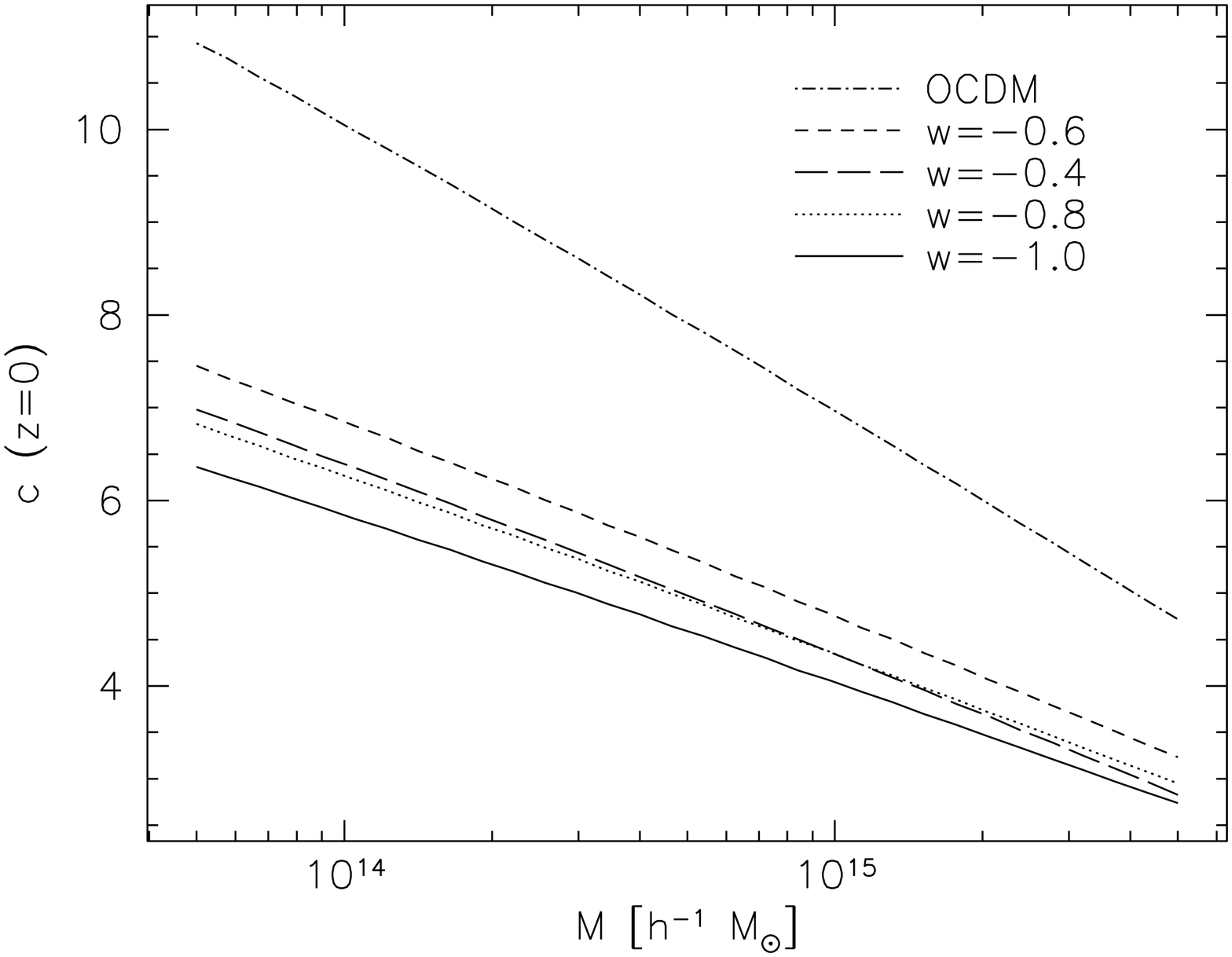}
\caption{Halo concentrations according to \citet{EK01.1} are shown as
  functions of halo mass for four spatially flat cosmological models
  with different choices for the dark energy parameter $w$ and for the
  OCDM model, as indicated. As $w$ increases, halos become more
  concentrated until $w\approx-0.6$. If $w$ increases further, halo
  concentrations drop because then the amplitude $\sigma_8$ of the
  power spectrum has to decrease rapidly in order to remain consistent
  with the COBE-DMR data, as the integrated Sachs-Wolfe effect becomes
  larger (cf.~\citealt{BA02.1}).}
\label{fig:1}
\end{figure}

We will adopt the latter method for this paper because the
concentrations computed from the algorithm by Navarro et al.~drop too
slowly with redshift compared to numerical simulations, and the
algorithm by Bullock et al.~has problems for high halo masses because
of the requirement that a fixed fraction of the final halo mass should
equal the nonlinear mass, which may never be reached if the halo mass
is high. Concentrations as a function of halo mass for four different
choices of the dark energy parameter $w$ are shown in
Fig.~\ref{fig:1}.

\subsection{Strong Lensing by NFW Haloes}

\begin{figure}[ht]
  \includegraphics[width=\hsize]{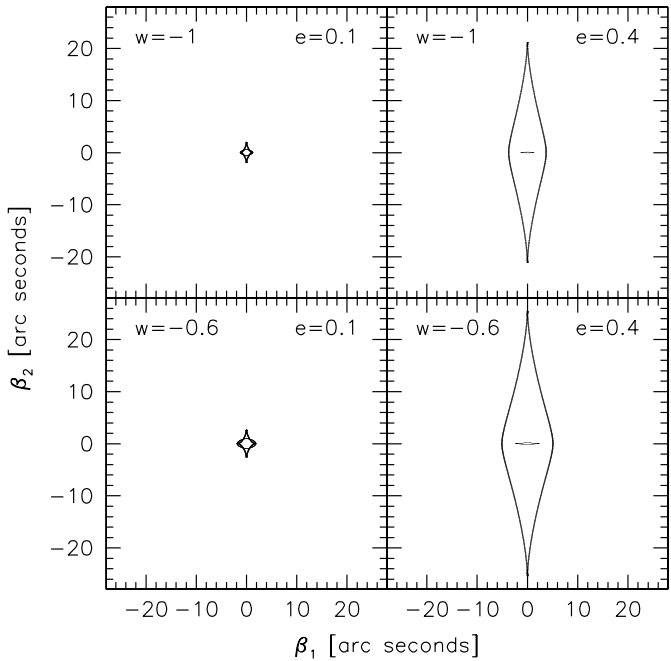}
\caption{Tangential and radial caustics for NFW lens models with
  elliptically distorted lensing potential. A halo of mass
  $10^{15}\,h^{-1}\,M_\odot$ at redshift $z=0.3$ is assumed here, and
  the sources are all placed at $z_\mathrm{s}=1$. The cosmological
  model is spatially flat with $\Omega_0=0.3$ and normalised to the
  COBE-DMR data. The upper panels show the caustics for a dark energy
  parameter of $w=-1$ and for two ellipticities, $e=0.1$ and
  $e=0.4$. The lower two panels show the caustics for $w=-0.6$. The
  figure illustrates the high sensitivity of strong lensing to halo
  ellipticity and concentration.}
\label{fig:2}
\end{figure}

The arc cross section of a galaxy cluster is the area in the source
plane where a source has to lie for being imaged as an arc with
specified properties, e.g.~exceeding a threshold length-to-width
ratio. A typical cluster lens has two critical curves, defined as
curves in the lens plane along which the Jacobian matrix of the lens
mapping is singular, and the image magnification is formally infinite
(cf.~Fig.~\ref{fig:2}). The critical curves are the images of the
caustic curves, thus sources close to a caustic are imaged as highly
magnified and distorted images. The two critical curves and their
corresponding caustic are called tangential and radial, because of the
dominant orientation of the image distortion relative to the centre of
the lens. Seeking to quantify large-arc cross sections, we are thus
looking for an appropriately defined area covering the tangential
caustic of a cluster.

It is well known that arc cross sections depend strongly on the exact
shape of the cluster mass distribution, and on the gravitational tidal
field exerted by density fluctuations in its neighbourhood
\citep{BA95.1}. Thus, the only reliable method for exactly quantifying
arc cross sections has to use numerically simulated cluster models
without referring to any symmetry assumptions. For a study like ours,
however, we only need to describe how arc cross sections are expected
to change relative to a fiducial model when certain cosmological
parameters are modified.

Gravitational lensing by an NFW halo can be described by its lensing
potential (e.g.~\citealt{BA96.1,ME03.1}),
\begin{equation}
  \psi(x)=4\kappa_\mathrm{s}\left[
    \frac{1}{2}\ln^2\frac{x}{2}-
    2\mathrm{arctanh}^2\sqrt\frac{1-x}{1+x}\right]\;,
\label{eq:2}
\end{equation}
which is related to the lensing convergence through the Poisson
equation
\begin{equation}
  \kappa(x)=\frac{1}{2}\nabla^2\psi(x)\;.
\label{eq:3}
\end{equation}
The factor $\kappa_\mathrm{s}$ in Eq.~(\ref{eq:2}) is defined by
\begin{equation}
  \kappa_\mathrm{s}=
  \frac{\rho_\mathrm{s}r_\mathrm{s}}{\Sigma_\mathrm{cr}}\;,
\label{eq:2a}
\end{equation}
where $\Sigma_\mathrm{cr}$ is the critical surface mass density for
lensing. The dimensionless radius $x=r/r_\mathrm{s}$ can conveniently
be replaced by the angular radius
$\theta=r/D_\mathrm{d}=xr_\mathrm{s}/D_\mathrm{d}$, where
$D_\mathrm{d}$ is the angular diameter distance from the observer to
the lens. The gravitational tidal field, or shear, of the lens is the
two-component quantity
\begin{equation}
  \gamma_1=\frac{1}{2}(\psi_{11}-\psi_{22})\;,\quad
  \gamma_2=\psi_{12}\;,
\label{eq:4}
\end{equation}
where the subscripts abbreviate partial derivatives with respect to
the angular coordinates $(\theta_1,\theta_2)$ on the sky. The
deflection angle is the gradient of the lensing potential,
\begin{equation}
  \vec\alpha(\theta)=\nabla\psi(\theta)\;.
\label{eq:5}
\end{equation}

The Jacobian matrix of the lens mapping has the components
\begin{equation}
  \mathcal{A}_{ij}=\delta_{ij}-\psi_{ij}\;.
\label{eq:6}
\end{equation}
Its eigenvalues are $\lambda_\pm=(1-\kappa)\pm\gamma$, where
$\gamma=(\gamma_1^2+\gamma_2^2)^{1/2}$ is the amplitude of the
shear. The tangential critical curve is determined by the condition
$\lambda_-=(1-\kappa)-\gamma=0$.

We now distort the axially symmetric NFW lens such that the
iso-potential lines become ellipses,
\begin{equation}
  \psi(\theta)\to\psi(\vartheta)\;,\quad
  \vartheta=\left[\frac{\theta_1^2}{1-e}+\theta_2^2(1-e)\right]^{1/2}\;.
\label{eq:7}
\end{equation}
As noted by \citet{KA93.1}, an elliptical potential can lead to
dumbbell-shaped mass distributions with locally negative mass
density. As real clusters are irregular, dumbbell-shaped mass
distributions are acceptable. For the NFW profile with elliptical
isopotential contours, the mass density does indeed become mildly
negative, but only well outside the core where strong lensing
occurs. For $e=0.4$, the minimum $\kappa$ is $\sim-0.01$ times the
convergence in the core (see also \citealt{GO02.1}). We thus use the
elliptical lensing potential (\ref{eq:7}) for computational
simplicity.

The Jacobian matrix and its eigenvalues can be computed from
(\ref{eq:7}) using the relations introduced before. Generally, the
zeroes of the tangential eigenvalue $\lambda_-$ have to be determined
numerically. On the coordinate axes, they are given by
\begin{equation}
  \vartheta=\left\{\begin{array}{ll}
    \displaystyle
    (1-e)\alpha        & (\theta_2=0) \\
    \displaystyle
    \frac{\alpha}{1-e} & (\theta_1=0) \\
		   \end{array}\right.\;.
\label{eq:8}
\end{equation}
The corresponding caustic points, i.e.~the cusps of the diamond-shaped
caustic on coordinate axes, can then be found using the lens equation,
\begin{equation}
  \vec\beta=\vec\theta-\vec\alpha(\vec\theta)\;.
\label{eq:9}
\end{equation}
We thus know the four intersection points of the tangential caustic
curve with the coordinate axes. We defined them to lie at
$(\theta_1,\theta_2)=(0,\pm a)$ and $(\theta_1,\theta_2)=(\pm b, 0)$.
Since the major axis of the iso-potential ellipses points along the
$\theta_2$ axis, the cusps on the $\theta_2$ axis are further away
from the lens centre than the cusps on the $\theta_1$ axis, hence
$a>b$. Figure~\ref{fig:2} illustrates the caustic curves for
elliptical NFW lens models with two different ellipticities in two
spatially flat cosmological models with different values for $w$.

A simple assumption for the large-arc cross section $\sigma$ of the
elliptically distorted NFW lens holds that it is proportional to the
area enclosed by the critical curve, hence $\sigma\propto ab$, with a
proportionality constant depending on the exact shape of the caustic
curve, and thus on the ellipticity of the lens model. Since we do not
require an absolute calibration of the arc cross sections, we adopt
$\sigma=ab$. Earlier work \citep{ME00.1,FL00.1} has shown that
individual galaxies have a negligible effect on arc cross sections,
which further supports the assumption that they are determined by the
overall extent of the caustic curves.

\begin{figure}[ht]
  \includegraphics[width=\hsize]{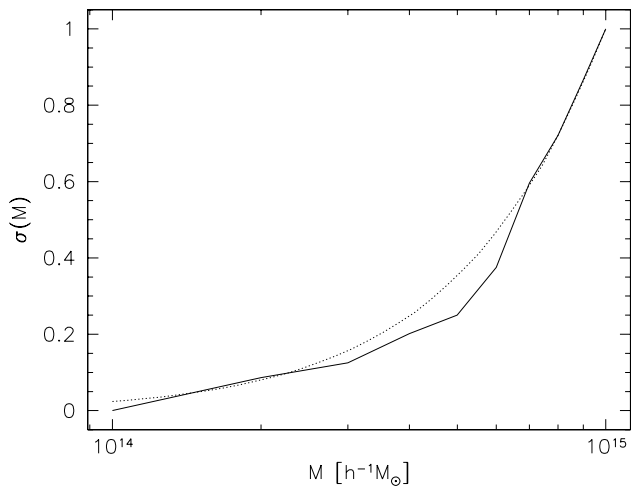}
\caption{Comparison between the numerically determined arc cross
  sections of elliptically distorted NFW lenses as a function of their
  virial mass $M$ (solid curve), and the simple estimate for the arc
  cross sections which is proportional to the area covered by the
  tangential caustic curve (dotted curve). Both curves are arbitrarily
  normalised to unity at the high-mass end. The lenses are placed at
  redshift $z=0.3$, the sources at redshift $z_\mathrm{s}=1.0$, the
  ellipticity of the NFW potential is $e=0.4$. The curves were
  computed for a dark energy cosmological model with
  $w=-0.6$. Evidently, the simple estimate for the cross sections
  correctly reproduces the scaling of the cross sections with cluster
  mass. The numerically determined curve is not smooth because the
  cross sections are computed from finite numbers of simulated arcs
  and thus subject to random fluctuations.}
\label{fig:5}
\end{figure}

We will later have to integrate over the cluster population,
conveniently parameterised by the virial mass and described by the
mass function. Thus, we have to verify whether our approximate
description of the arc cross section scales with cluster mass in the
same way as numerically determined arc cross sections of the same
cluster models, given identical ellipticity parameters $e$. We
therefore set up deflection-angle maps starting from the elliptically
distorted lensing potential $\psi(\vartheta)$ and used them for
imaging randomly distributed, intrinsically elliptical sources by
tracing rays passing the image plane on a rectangular grid. The
sources are placed on adaptively refined grids whose resolution is
progressively increased in the vicinity of caustic curves. The images
are automatically classified according to their length, width,
length-to-width ratio and several other parameters, and the cross
section is determined by counting the number of images exceeding a
threshold length-to-width ratio. More detail on this method can be
found in \citep{BA94.1,ME00.1}. Results are shown in Fig.~\ref{fig:5}.

The essential feature of the numerically determined arc cross sections
is that they increase approximately quadratically with the lens
mass. Since the Einstein radius of an extended lens typically scales
linearly with the lens mass, this indicates that the arc cross section
scales approximately like the square of the Einstein radius. Our
simple estimate for the arc cross section is defined to reproduce this
property.

\begin{figure}[ht]
  \includegraphics[width=\hsize]{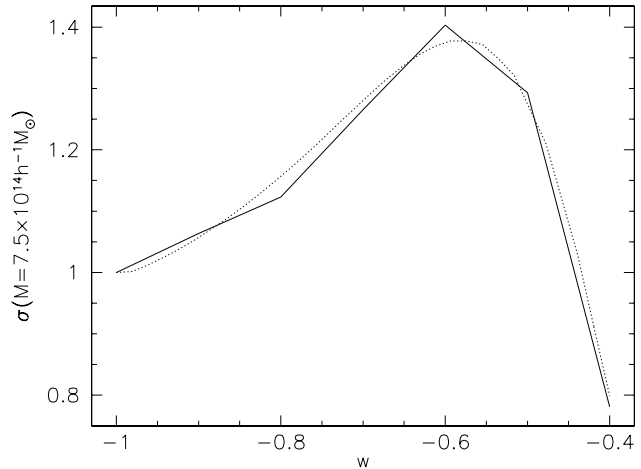}
\caption{Comparison between cross sections similar to
  Fig.~\ref{fig:5}, but for a fixed halo mass of
  $M=7.5\times10^{14}\,h^{-1}M_\odot$ and varying dark energy
  parameter $w$. The solid curve showing the numerically determined
  cross sections closely follows the dotted curve, which represents
  the simple cross-section estimate introduced here. The curves are
  arbitrarily normalised to unity at their starting point, i.e.~at
  $w=-1$. Increasing $w$ from $-1$ to $-0.6$ increases the cross
  sections by a factor of $\sim1.9$. As in Fig.~\ref{fig:5}, lens and
  source redshifts are set to $0.3$ and $1$, respectively, and the
  ellipticity of the NFW lensing potential is set to $e=0.2$.}
\label{fig:6}
\end{figure}

We also check whether numerically determined cross sections scale with
the dark energy parameter $w$ in a similar way as our simple
cross-section estimate does. Figure~\ref{fig:6} shows the example of
an elliptically distorted NFW lens of mass
$M=7.5\times10^{14}\,h^{-1}M_\odot$. The curves plotted there show
qualitatively the same behaviour. As $w$ increases above $-1$, the arc
cross section increases by a factor of $\sim1.9$ until $w\approx-0.6$,
and then drops as the normalisation constraint requires to
significantly reduce $\sigma_8$. Both curves are arbitrarily
normalised to unity at their peaks. The differences between the
numerically determined cross sections and our simple estimate are
unlikely to be significant because the numerical method uses the
images of a finite number of discrete sources for determining the
cross sections, which causes noise in the results. In any case, the
two essential properties of the change in the cross sections with $w$
are well reproduced, namely the amplitude of the relative increase and
the location of the peak.

Finally, we set the cross section to zero if the major axis of the
critical curve falls below some threshold $\theta_\mathrm{min}$. The
idea behind this is that if the critical curves become too small, the
images near the critical curves can hardly be called giant
arcs. Suppose typical sources have diameters on the order of an arc
second, and the lens should be able to produce arcs with a
length-to-width ratio around ten. Then, ignoring the source
magnification in the radial direction, the tangential critical curve
needs to have a radius of approximately $10''$ for this to happen. We
thus set $\theta_\mathrm{min}=10''$ unless stated otherwise, and show
the effect of changing $\theta_\mathrm{min}$ to $5''$ below.

We conclude from this section that the scaling of our simple estimate
for arc cross sections with lens mass and with the dark energy
parameter $w$ well reproduces what is expected from numerical
treatments of the same lens models, i.e.~NFW lenses with elliptically
distorted lensing potential. We emphasise again that the absolute
value of the cross sections are unimportant for our present purposes,
as we are aiming at studying the change in the arc-formation
probability in various cosmological models \emph{relative} to the
$\Lambda$CDM model.

\section{Arc Probabilities}

We can now proceed to compute the probability for arc formation by a
population of clusters. Given a mass function $\d n(M,z)/\d M$, we can
write the so-called optical depth $\tau(z_\mathrm{s})$ as
\begin{equation}
  \tau(z_\mathrm{s})=\int_0^{z_\mathrm{s}}\d z\,
  (1+z)^3\left|\frac{\d V}{\d z}\right|\,
  \int_{M_\mathrm{min}}^\infty\d M\,
  \frac{\d n}{\d M}\,\sigma(M,z)\;,
\label{eq:P1}
\end{equation}
where $z_\mathrm{s}$ is the source redshift, $V$ is the cosmic volume,
and the factor $(1+z)^3$ accounts for the fact that the mass function
is defined per comoving volume. The lower mass limit $M_\mathrm{min}$
is determined by the mass required to produce critical curves whose
major axis exceeds the threshold $\theta_\mathrm{min}$ introduced in
the preceding section, thus it depends on the lens redshift $z$.

For the mass function $\d n/\d M$, we choose the modification by
\citet{SH99.1} of the \citet{PR74.1} mass function. It well reproduces
the halo mass functions found in numerical simulations. We take into
account that our definition of mass differs slightly from Sheth \&
Tormen's in that we use the mass enclosed by a sphere in which the
mean density is 200 times the \emph{critical} rather than the
\emph{mean} density. The mass function depends on cosmology through
the normalisation of the power spectrum and the linear overdensity
parameter derived from the spherical collapse model.

We note an important difference to strong lensing by galaxies. While
the population of galaxy lenses is well described as isothermal
spheres with number counts derived from observations
(e.g.~\citealt{LE00.1,KE02.1}), the lack of wide-separation lenses
argues against isothermal density profiles in cluster-scale lenses
\citep{FL94.1,PO00.1}. Baryonic physics changes the central density
profiles of galaxy-scale halos, but is inefficient on cluster scales
(e.g.~\citealt{KO01.1}). Cluster mass functions derived from X-ray
observations are found to agree well with theoretical predictions
based on Press-Schechter type models (e.g.~\citealt{RE02.1}). Thus
modelling the cluster population with NFW density profiles and the
Sheth-Tormen mass functions is well justified.

\begin{figure}[ht]
  \includegraphics[width=\hsize]{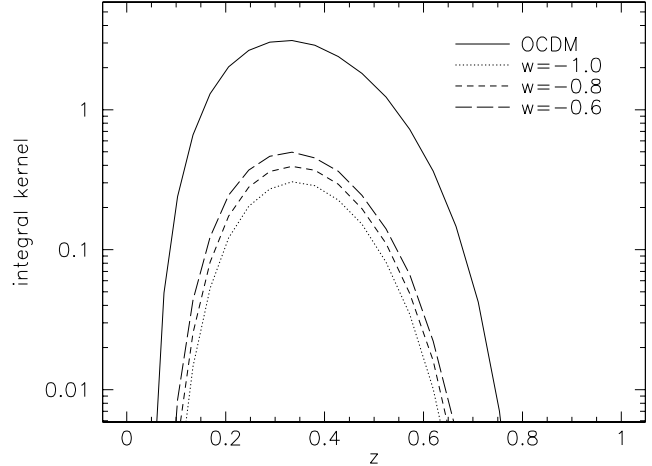}
\caption{The redshift integrand of Eq.~(\ref{eq:P1}) is plotted as a
  function of $z$ for four different cosmological models, a
  $\Lambda$CDM model with $\Omega_0=0.3$, and open CDM model with
  $\Omega_0=0.3$ and no cosmological constant, and two spatially-flat
  dark energy models with $\Omega_0=0.3$ and $w=-0.8$ and
  $w=-0.6$. The spatially-flat models are normalised to the COBE-DMR
  data, and the open CDM model has the same $\sigma_8$ as the
  cosmological-constant model for easier comparison. The curves show
  that our simple analytic model succeeds in reproducing the
  order-of-magnitude difference between the open and the
  cosmological-constant model found in numerical simulations, and that
  spatially flat dark energy models cannot bridge the gap between
  these two models. The ellipticity of the NFW lensing potential was
  set to $e=0.3$ here.}
\label{fig:3}
\end{figure}

We show in Fig.~\ref{fig:3} the redshift integrand of
Eq.~(\ref{eq:P1}), i.e.~the cosmic volume times the integral over mass
of the mass function times the arc cross section. The figure thus
illustrates the total arc cross section contributed by the cluster
population at redshift $z$. For simplicity, sources are assumed to be
at a single redshift of $z_\mathrm{s}=1$ here. The four curves in
Fig.~\ref{fig:3} are for the $\Lambda$CDM model with $\Omega_0=0.3$,
the open CDM model with the same $\Omega_0$ but with
$\Omega_\Lambda=0$, and for dark energy models with the same
$\Omega_0$ but $\Omega_\mathrm{Q}=0.7$, and $w=-0.6$ or
$w=-0.8$. Again, the curves are arbitrarily normalised such that the
optical depth $\tau$ of the OCDM model is unity. The value of
$\sigma_8=0.88$ of the $\Lambda$CDM model was also set for the OCDM
model for easier comparison. The COBE-DMR data would require
$\sigma_8\sim0.4$ for the OCDM model, which is way below the value
required for reproducing the observed number density of massive
clusters. We thus have to choose $\sigma_8$ for the OCDM model in
conflict with the COBE-DMR data in order to produce comparable results
on arc statistics.

Figure~\ref{fig:3} shows two important results. First, the simple
model for arc cross sections introduced here is capable of reproducing
the order-of-magnitude difference in the total arc cross section
between the $\Lambda$CDM and the OCDM models that had been found
earlier in numerical simulations, and could not be reproduced by
analytic models based on singular isothermal cluster mass
distributions. Second, although the dark energy models have a somewhat
higher total arc cross section than the $\Lambda$CDM model, they are
still by a factor of $\sim6$ below the arc cross section for the OCDM
model. According to our analytic estimates, dark energy models are
thus unable to reconcile spatially flat cosmological models with low
matter density with the high abundance of large arcs, which seems to
be similarly impossible with cosmological constant models given the
earlier numerical results.

\begin{figure}[ht]
  \includegraphics[width=\hsize]{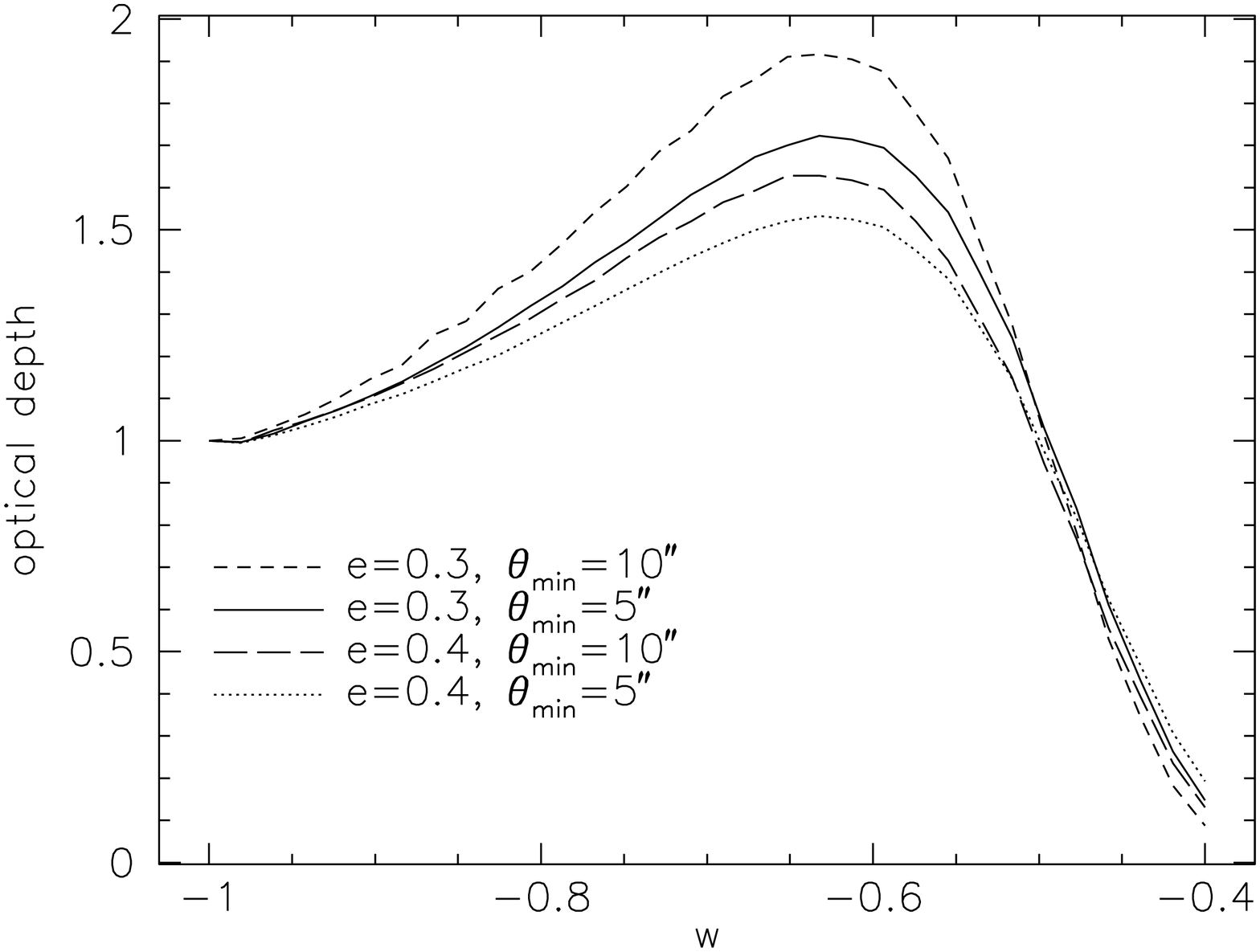}
\caption{Optical depth $\tau(z_\mathrm{s}=1)$ defined by
  Eq.~(\ref{eq:P1}) for spatially-flat dark energy models as functions
  of $w$. All curves are arbitrarily normalised to their values for
  $w=-1$, i.e.~they show the change in optical depth with $w$ relative
  to the cosmological-constant model. All curves were obtained for
  $\Omega_0=0.3$ and COBE-normalised CDM power spectra. Results are
  shown for two different choices each for the ellipticity $e$ and the
  cutoff angle $\theta_\mathrm{min}$, as indicated. The curves show
  that clusters in dark energy models can be $\sim50\%$ up to
  $\sim100\%$ more efficient in forming arcs than in
  cosmological-constant models, depending in detail on the exact
  choices for the ellipticity and the cutoff angle.}
\label{fig:4}
\end{figure}

Figure~\ref{fig:4} shows the optical depth $\tau$ as defined in
Eq.~(\ref{eq:P1}) for four different choices of the free parameters we
have introduced, namely the ellipticity $e$ of the lenses and the
cutoff radius $\theta_\mathrm{min}$. All curves have in common that
the arc optical depth increases noticeably as $w$ increases from $-1$
to $\sim-0.6$, and drops rapidly as $w$ is increased further. The
curves are intended to show the relative increase in $\tau$ compared
to the $\Lambda$CDM model and are thus normalised to unity at
$w=-1$. The peak amplitudes range from $1.5$ to $2$. The largest
increase is achieved for $\theta_\mathrm{min}=10''$ and $e=0.3$, the
smallest for $\theta_\mathrm{min}=5''$ and $e=0.4$.

Several effects act jointly here. First, we saw in Fig.~\ref{fig:6}
that the arc cross section of (one example for) an individual halo
increases by a factor of $\lesssim2$ as $w$ is increased from $-1$ to
$-0.6$. An additional effect is that, as the halo concentration
increases, halos of lower mass become capable of strong
lensing. Since the mass function of galaxy clusters is steep, a small
extension of the mass range towards lower masses can markedly increase
the number of clusters available for strong lensing, but the
requirement that arcs should be large imposes a lower limit on the
cluster masses. We see the combined effect in Figs.~\ref{fig:3} and
\ref{fig:4}. If we set the cutoff radius to
$\theta_\mathrm{min}=10''$, we select for higher-mass clusters in the
first place, whose mass function is steeper than for lower-mass
clusters. Thus, the effect of lowering the lower mass limit by
increasing the halo concentrations is more pronounced if the cutoff
radius is chosen higher.

Increasing the ellipticity of the lensing potential increases its
gravitational tidal field, or shear. The increase in halo
concentration caused by the earlier halo formation in dark energy
models with $w>-1$ is then relatively less important for arc
formation, which explains why the arc optical depth caused by cluster
populations with lower $e$ changes more strongly with $w$ than for
clusters with higher $e$. Fitting the elliptical NFW model to
numerically simulated clusters yields values for $e$ closer to $0.3$.

\section{Summary and Conclusions}

We investigated how the probability for the formation of large
gravitational arcs in galaxy clusters is expected to change as the
underlying cosmological model is modified. The main reason for this
investigation was our earlier finding that halo concentrations in a
simple class of spatially flat, dark-energy dominated cosmological
models are expected to depend sensitively on the equation-of-state
parameter $w$ \citep{BA02.1}. Strong lensing in general, and the
formation of large arcs in particular, is a highly nonlinear effect
which depends sensitively on the matter concentration in the lens
cores. It could therefore reasonably be expected that arc
probabilities would change significantly with $w$.

This is of cosmological relevance since earlier numerical simulations
\citep{BA98.2} showed that clusters in cosmological-constant models
fall short by about an order of magnitude of reproducing the abundance
of observed large gravitational arcs. While virtually all recent
cosmological experiments favour a spatially flat, low-density
universe, arc statistics apparently strongly prefers a low-density,
open model over a low-density model with cosmological constant. This
numerical result was questioned on the basis of simple analytic models
which failed to reproduce the strong dependence of arc statistics on
$\Omega_\Lambda$ found in the simulations \citep{CO99.2,KA00.1}. Our
first goal was thus to investigate whether the numerical results could
be supported by a more detailed analytic model.

Earlier studies also showed that analytic models for arc cross
sections are notoriously problematic and inadequate for an accurate
absolute quantification of arc probabilities. However, since it was
our main interest in this paper to quantify \emph{changes} in the arc
optical depth \emph{relative} to cosmological-constant models, we
applied reasonably flexible analytic lens models which we demonstrate
to reproduce the relevant features of fully numerical results
(e.g.~\citealt{BA98.2,ME03.1}), as described below.

We have chosen to use lenses with NFW density profile whose lensing
potential was elliptically distorted. This model has several
advantages. First, it naturally incorporates the dependence of central
halo concentration on cosmology, through the halo formation
time. Second, it agrees with density profiles consistently found in
numerical simulations. Third, the elliptical distortion strongly
decreases the mismatch between analytic and fully numerical lens
models \citep{ME03.1}. Although elliptical lensing potentials can lead
to locally negative surface mass densities for sufficiently large
ellipticities, this is a mild effect which happens only well outside
the core for our model, thus it should be irrelevant for our
purposes. We further adopted a simple estimate for the arc cross
section and verified that this estimate scaled with cluster mass and
dark energy parameter $w$ in the same way as fully numerically
determined arc cross sections do. In particular, our simple analytic
description for the arc cross section well reproduces the
approximately quadratic scaling with cluster mass obtained fully
numerically from the same elliptically distorted NFW cluster mass
models. Although we would not trust this model for any accurate
quantitative prediction of arc probabilities, we are confident that it
can be used for the \emph{relative} statements intended here.

Our main findings are as follows:
\begin{itemize}
\item Our simple lens model is indeed capable of reproducing the
  order-of-magnitude difference between $\Lambda$CDM and open CDM
  models found in the earlier numerical study. This shows that the
  change in halo concentration between the two models can explain the
  sensitivity of arc statistics for the cosmological constant. An
  additional effect into the same direction is contributed by the
  steep mass function of galaxy clusters. An increase in halo
  concentration lowers the minimum mass required for significant
  strong lensing, and this makes many more halos available for arc
  formation.
\item Although increasing the dark energy parameter $w$ has a
  noticeable effect on the optical depth for arc formation, it cannot
  increase the arc optical depth to a level compatible with that found
  in open CDM models. This result arises due to a combination of three
  main effects: first, individual halos of fixed mass get more
  concentrated in dark energy than in cosmological constant models;
  second, lower mass halos than before become able to form large
  arcs; and third, the requirement that arcs be large imposes a lower
  limit on cluster masses. Our analytic model thus suggests that arc
  statistics cannot be reconciled with low-density, spatially flat
  cosmological models which are now dominated by dark energy, i.e.~the
  discrepancy between arc statistics and the cosmological model
  favoured by most, if not all, recent cosmological experiments is not
  expected to disappear if the dark energy is not a cosmological
  constant, but has an equation of state $p=w\rho c^2$ with $w>-1$.
\end{itemize}

Being based on several simple analytic estimates, this study can only
provide a tentative answer. Detailed numerical simulations will be
necessary for reliable absolute quantifications of the arc optical
depth expected in cosmological models with dark energy instead of a
cosmological constant; similarly, it is necessary to quantify the
dependence of the effect we find here on the specific dark energy
model considered, such as a quintessence scalar field.

According to our preliminary analytic results presented here, it
appears that the solution to the arc statistics problem as described
by \cite{BA98.2} is probably not to be found in the cosmological model
alone, but more in the details of cluster structure and the history of
cluster evolution. The recent suggestion by \cite{WA03.1} that it
could simply be removed by considering higher-redshift sources is
certainly not a viable solution because the problem arose from
comparing simulations to a well-defined sample of highly X-ray
luminous clusters which were observed for arcs down to a well-defined
photometric limit \citep{LE94.1,LU99.1}. In combination with other
recent results on the strong-lensing properties of dark-matter halos
(e.g.~\citealt{OG01.1,OG02.1,OG03.1}) and the puzzling discoveries of
very high-redshift arcs in high-redshift clusters
(\citealt{ZA03.1,GL03.1}), this seems to imply that arc statistics are
teaching us that the properties of individual clusters and their
evolution over time is still insufficiently understood.

\acknowledgements{We wish to thank Peter Schneider, Achim Weiss and
  Simon White for useful discussions and comments. This work has been
  partially supported by Italian MIUR (Grant 2001, prot. 2001028932,
  ``Clusters and groups of galaxies: the interplay of dark and
  baryonic matter''), CNR and ASI.}

\bibliography{../TeXMacro/master}
\bibliographystyle{../TeXMacro/aa}

\end{document}